\documentclass[]{aa}
\usepackage{graphicx}
\usepackage{natbib}
\usepackage{longtable}
\usepackage[english]{babel}
\usepackage{url}
\usepackage{txfonts}
\usepackage{color}

\begin{document}
   \title{
          VLBA observations of
          a rare multiple quasar imaging event caused by refraction in the interstellar medium}
   \titlerunning{Refractive multiple imaging of the quasar 2023+335}
%   \subtitle{}

   \author{A.~B.~Pushkarev
          \inst{1,2,3}
          \and
          Y.~Y.~Kovalev\inst{4,1}
	  \and
	  M.~L.~Lister\inst{5}
	  \and
	  T.~Hovatta\inst{6}
	  \and
	  T.~Savolainen\inst{1}
	  \and 
	  M.~F.~Aller\inst{7}
	  \and
	  H.~D.~Aller\inst{7}
	  \and
          E.~Ros\inst{8,9,1}
          \and
	  J.~A.~Zensus\inst{1}
	  \and
	  J.~L.~Richards\inst{5}
	  \and
	  W.~Max-Moerbeck\inst{6}
	  \and
	  A.~C.~S.~Readhead\inst{6}
	  %\fnmsep
%	  \thanks{{\it Send offprint request to}: A. B. Pushkarev}
          }

   \institute{Max-Planck-Institut f\"ur Radioastronomie, Auf dem H\"ugel 69, 53121 Bonn, Germany\\
              \email{apushkar@mpifr.de}
         \and 
             Pulkovo Astronomical Observatory, Pulkovskoe Chaussee 65/1, 196140 St. Petersburg, Russia
	 \and
             Crimean Astrophysical Observatory, 98688 Nauchny, Crimea, Ukraine
         \and
             Astro Space Center of Lebedev Physical Institute, Profsoyuznaya 84/32, 117997 Moscow, Russia
         \and
             Department of Physics, Purdue University, 525 Northwestern Avenue, West Lafayette, IN 47907, USA
         \and 
             Cahill Center for Astronomy \& Astrophysics, California Institute of Technology, 1200 E. California Blvd, Pasadena, CA 91125, USA
         \and
             Department of Astronomy, University of Michigan, 830 Dennison Building, Ann Arbor, MI 48109-1042, USA
         \and 
             Observatori Astron\`omic, Universitat de Val\`encia, Apartat de  Correus 22085, E-46071 Val\`encia, Spain
         \and
             Departament d'Astronomia i Astrof\'{\i}sica, Universitat de Val\`encia, Dr. Moliner 50, E-46100 Burjassot, Val\`encia, Spain
             %\thanks{}
             }

   \date{Received 15 March 2013; accepted 22 May 2013}
 
\abstract
 %  Context heading (optional)
 {}
 % Aims heading (mandatory)
 {We have investigated highly atypical morphological parsec-scale
   changes in the flat spectrum extragalactic radio source 2023+335
   which are coincident with an extreme scattering event (ESE) seen at
   radio wavelengths during the first half of 2009. }
 % Methods heading (mandatory)
 {We used 
  (i) 15.4~GHz Very Long Baseline Array (VLBA) observations of the quasar 2023+335 
  obtained at 14 epochs between July 2008 and Nov. 2012 as part of the Monitoring 
  Of Jets in Active galactic nuclei with VLBA Experiments (MOJAVE) program, (ii) 
  earlier archival VLBA observations of the source performed at 1.4, 2, 8, 15, 22, 
  and 86~GHz to analyze the properties of the proposed turbulent screen toward 
  2023+335, and (iii) data sets from the Owens Valley Radio Observatory (OVRO) and 
  University of Michigan Radio Astronomy Observatory (UMRAO) single-dish monitoring 
  programs performed at 15 and 14.5~GHz, respectively, to study integrated flux 
  density changes.
 }
 % Results heading (mandatory)
 {We report on the first detection of the theoretically-predicted rare
   phenomenon of multiple parsec-scale imaging of an active galactic
   nucleus induced by refractive effects due to localized foreground
   electron density enhancements, e.g., in an AU-scale plasma lens(es)
   in the ionized component of the Galactic interstellar medium. We
   detected multiple imaging in the low galactic latitude
   ($b=-2\fdg4$) quasar 2023+335 from the 15.4~GHz MOJAVE observations
   when the source was undergoing an ESE.  While the parsec-scale jet
   of the source normally extends along PA\,$\sim-20\degr$, in the 28
   May 2009 and 23 July 2009 images a highly significant
   multi-component pattern of secondary images is stretched out nearly
   along the constant galactic latitude line with a local
   PA\,$\approx40\degr$, indicating that the direction of relative
   motion of the plasma lens is close to orbital. Weaker but still
   detectable imaging patterns at similar position angles are
   sporadically manifest at several other epochs.  Modeling the ESE
   that occurred in early 2009 and lasted $\sim$0.14~yr, we determined
   that the foreground screen has a double-lens structure, with proper
   motion ($\sim6.8$~mas\,yr$^{-1}$), and angular size
   ($\sim0.27$~mas). We also found that the angular separation between
   the two brightest sub-images roughly follows a wavelength-squared
   dependence expected from plasma scattering.  Furthermore, by
   analyzing archival non-simultaneous VLBA observations covering
   a wide frequency range from 1.4 to 86~GHz, we found that the
   scattered angular size of the VLBI core follows a $\nu^{-1.89}$
   dependence, implying the presence of a turbulent, refractive
   dominated scattering screen that has a confined structure or is
   truncated transverse to the line of sight toward 2023+335.  }
 % Conclusions heading (optional)
 {}

   \keywords{galaxies: active --
             galaxies: jets --
	     quasars: individual: 2023+335 --
	     turbulence --
	     scattering
	     }

   \maketitle
%
%________________________________________________________________

\section{Introduction}

Radio waves are influenced by propagation effects whenever passing
through an ionized medium containing free-electron density
fluctuations, e.g., in the Earth's ionosphere, the interplanetary
medium, or the interstellar medium (ISM). These effects are especially
prominent in observations of compact bright sources, such as pulsars,
masers, and active galactic nuclei (AGN). They are represented by a
wide variety of scattering signatures, including angular broadening,
ionospheric or interstellar scintillations, and rare phenomena in AGN
radio light curves often referred to as extreme scattering
events \citep[ESEs;][]{Fiedler87}.

The gaseous interstellar medium is characterized by a large Reynolds
number, reflecting its highly turbulent nature. The ISM is structured
as a hierarchy of clouds that appears self-similar over six orders of
magnitude in linear scale \citep{Combes00}. The turbulence in the ISM
is mainly driven by supernova explosions, spiral arm instabilities,
stellar winds, and cosmic rays \citep{Elmegreen04,Lazarian06}.
Small-scale turbulent electron density fluctuations can give rise to
diffractive scattering, manifested as angular broadening of compact
radio sources, while large-scale fluctuations can produce refractive
scattering that modulates the overall flux density level, as in the
case of scintillations or ESEs. The relative importance of refractive
and diffractive scattering depends strongly on the form of the spatial
power spectrum of electron density turbulence 
\citep{Goodman85,Cordes86b,Armstrong95,Chepurnov10}.

The ESE phenomenon, which is characterized by dramatic
frequency-dependent changes in source flux density, was first detected
during long-term 2.7 and 8.1~GHz monitoring of compact extragalactic
radio sources \citep{Fiedler87}.  Because of the simultaneity of the
events at different frequencies and speed-of-light travel time
arguments, \cite{Fiedler87} concluded that the events could not be
explained by intrinsic variability. It is now broadly accepted
\citep{Fiedler87,Fiedler94,Romani87,Clegg88,Clegg98} that ESEs can be
explained by strong scattering, and they occur when ionized material
with electron density enhancements and a transverse dimension of
$\sim$~AU passes in front of a distant background radio source.
Typically, such events last for several weeks to months, and they are
quite rare. To date, less than twenty ESEs have been confirmed.  What
is still unclear about ESEs is the physical nature of the ionized
structures (the ``lenses'') and their relationship to the different
phases of the interstellar medium. In this regard, several models have
been considered. \cite{Clegg88,Clegg98} have proposed that such lenses
can be associated with relatively isolated discrete structures in the
ISM.  Alternatively, \cite{Heiles97} discussed the possibility of a
low level cosmic ray ionization within a neutral structure.
\cite{Walker98} suggested a model of photo-ionized molecular clouds in
the Galactic halo. Studying HI absorption before and during an ESE,
\cite{Lazio01} found no changes in equivalent width, maximum optical
depth, and velocity at the maximum optical depth, but could not
completely rule out the molecular cloud model of \cite{Walker98},
since the observed velocity range covered only about 25\% of the
allowed range.  The typical electron densities
$n_e\gtrsim10^2$~cm$^{-3}$ inferred from modeling ESE light curves
\citep{Romani87,Clegg98} lead to pressures
$n_eT\gtrsim10^6$~K~cm$^{-3}$ that are $\sim10^3$ times larger than
the typical pressure in the ISM \citep{Kulkarni87}. To overcome this
problem, \cite{Romani87} suggested that ionization fronts and/or
cooling instabilities associated with old supernova remnants are the
possible sites of the lenses. Moreover, analyzing the high-resolution
spectra in the HST archive and deriving ISM pressures along the line 
of sight, \cite{Jenkins07} found that a large fraction of the ISM is 
at the canonical pressure, about $10^3$~K~cm$^{-3}$. However, they 
found a tail in the distribution, extending to pressures
$\gtrsim10^5$~K~cm$^{-3}$. Apparently, there are extremely over-pressure 
regions in the ISM. Their volume filing factor is low, and they may 
be short-lived, but they clearly exist.

One of the most intriguing predictions of the scattering theory is 
the possible creation of multiple images of a compact radio source 
seen through the turbulent screen with refractive dominated properties
\citep{Lovelace70,Cordes86b,Rickett88}. \cite{Goodman85} argued that
the image may be broken up into a small number of sub-images, each of
which is further fragmented by phase fluctuations on the next smaller 
scale and so on, thus forming a hierarchical structure similar to a 
fractal geometry. By investigating refraction in Gaussian plasma
lenses as a model of ESE, \cite{Clegg98} predicted the formation of
up to 3 images during periods when the caustic surfaces are formed.
\cite{Cordes01} also reported on a possible creation of multiple 
imaging by multiple descrete scattering screens. The observable 
effect of multiple imaging has been detected in the dynamic spectra 
of pulsars \citep{Cordes86,Cordes86b}. For AGN, however, there has 
yet to be a direct detection of the effect.  This is largely due to 
a very low occurrence rate of ESEs ($\sim 0.01$ event-years per 
source-year; \citealt{Fiedler94}), and the roughly twice as 
infrequent caustic surface periods when multiple imaging could 
potentially be detected. The only earlier set of VLBI observations 
of an AGN during an ESE (the quasar 1741$-$038) did not reveal any 
evidence of multiple imaging, presumably because of the
insufficient refractive strength of the lens \citep{Lazio00}.

In this paper, we describe the discovery of multiple imaging of an AGN
on parsec scales caused by refraction effects in localized structures
with electron density enhancements in the interstellar medium. We
analyze a sequence of the 15.4~GHz maps of the quasar 2023+335
obtained within the framework of the MOJAVE (Monitoring of Jets in
Active galactic nuclei with VLBA Experiments) program~\citep{MOJAVE}.
By modeling the dramatic changes in the total flux density attributed
to an extreme scattering event registered by the 15~GHz OVRO
observations near 2009.3, we constrain the basic physical
characteristics of the intervening plasma lens. Using available
multi-frequency VLBA  observations, we also study the angular
broadening of \object{2023+335} to derive the properties of the scattering
screen toward the source.

Throughout the paper, we use the $\Lambda$CDM cosmological model with
$H_0=71$~km~s$^{-1}$~Mpc$^{-1}$, $\Omega_m=0.27$, and
$\Omega_\Lambda=0.73$ \citep{Komatsu09}. All position angles are given
in degrees from north through east. The spectral index $\alpha$ is
defined according to the convention $S\propto\nu^\alpha$, where $S$ is
the flux density and $\nu$ the observing frequency.

\section{Observations and data reduction}
\label{s:obs}

The extragalactic radio source 2023+335 (\object{J2025+3343}) is a
flat-spectrum ($\alpha\approx+0.07$) quasar located at $z=0.22$
\citep{Sowards_Emmerd03}. The quasar is also detected at high
energies. It is positionally associated with the X-ray source 
\object{WGA J2025.1+3342} \citep{Sguera04} and with the $\gamma$-ray source 
\object{1FGL J2027.6+3335} detected by the {\it Fermi}-LAT \citep{Kara12}.
\subsection{VLBA observations}

2023+335 was included into the extension of the statistically complete
flux-density-limited MOJAVE sample \citep{MOJAVE} in July 2008 and has
since been monitored at approximately 4 month intervals with the VLBA
at an observing frequency of
15.4~GHz\footnote{\url{http://www.physics.purdue.edu/astro/MOJAVE}}.
The observations were made in dual circular polarization mode, and
recorded with a bit rate of 256~Mbps, which was increased to 512 Mbps
in the epoch 2008 October 3 and epochs thereafter.

The initial calibration was performed with the NRAO Astronomical Image
Processing System (AIPS) \citep{aips} following standard data
reduction techniques.  CLEANing \citep{CLEANref}, phase and amplitude
self-calibration \citep{Jennison58,Twiss_etal60} were performed in the
Caltech Difmap \citep{difmap} package. In all cases a point-source
model was used as an initial model for the iterative procedure. Final
maps were produced by applying natural weighting of the visibility
sampling function. The typical uncertainty of the obtained flux
densities is $\sim 5\%$. The absolute calibration of electric vector
position angles (EVPAs) is approximately $3\degr$ \citep{Hovatta_RM}. The
source structure was model-fitted in the visibility ($u,v$)
plane in Difmap using circular and elliptical Gaussian components. For
a more detailed discussion of the data reduction and imaging process
schemes, see \cite{MOJAVE,MOJAVE_VI}.

\subsection{Single-dish monitoring}

The quasar 2023+335 has been frequently monitored with single-dish
radio telescopes, including the Owens Valley Radio Observatory (OVRO)
40-m telescope at 15~GHz \citep{Richards11}. This program, which
commenced in late 2007, now includes about 1700 sources, each observed
with a nominal twice per week cadence. The OVRO 40~m uses off-axis
dual-beam optics and a cryogenic high electron mobility transistor
(HEMT) low-noise amplifier with a 15.0~GHz center frequency and 3~GHz
bandwidth. The two sky beams are Dicke-switched using the off-source
beam as a reference, and the source is alternated between the two
beams in an ON-ON fashion to remove atmospheric and ground
contamination. Calibration is achieved using a temperature-stable
diode noise source to remove receiver gain drifts, and the flux
density scale is derived from observations of 3C~286 assuming the
\cite{Baars77} value of 3.44~Jy at 15.0~GHz. The systematic
uncertainty in the flux density scale is estimated to be $\sim$5\%.
Complete details of the reduction and calibration procedure are given
by \cite{Richards11}.

Total flux density and linear polarization observations of 2023+335
were also obtained with the UMRAO Michigan 26-m paraboloid dish as
part of the Michigan extragalactic variable source-monitoring program
\citep{Aller85}. The polarimeter consists of dual rotating, linearly
polarized feed horns, which are symmetrically placed around the
paraboloid's prime focus, and feed a broadband uncooled HEMT
amplifier with central frequency 14.5~GHz and bandwidth 1.7~GHz.
Each daily observation of the target source is an average of 16
measurements over a 40-minute time period. The adopted flux density
scale is based on \cite{Baars77}. In addition to the observations of
this primary standard, observations of nearby secondary flux density
calibrators were interleaved with the observations of the target
source every 1.5 to 2 hours to verify the stability of the antenna
gain and to verify the telescope pointing. The EVPAs were calibrated
using a source of known polarized emission mounted at the vertex of
the paraboloid. To verify the calibration of the instrumental
polarization, selected galactic H~II regions were observed several
times each day.

\subsection{Fermi LAT observations}

The {\it Fermi} LAT light curve was calculated using an un-binned
likelihood analysis (tool gtlike), as implemented in the
ScienceTools-v9r27p1 package with P7V6 Source event selection. A
region of interest of $10\degr$ around the target position was used.
All sources within $15\degr$ of the target were included in the
likelihood model, and their spectral parameters were frozen to the
values obtained from the 2FGL catalog. Other selections and cuts were
based on the recommendations from the LAT Science team. We used
monthly binning and calculated an upper limit if the test statistic
(TS) value was 10 or less. During the flaring period
(2009.27--2009.65) we used weekly binning to resolve the temporal
substructure of the flare.

\section{Arguments for refractive scattering}
\label{s:refrac_scat}

\begin{figure*}
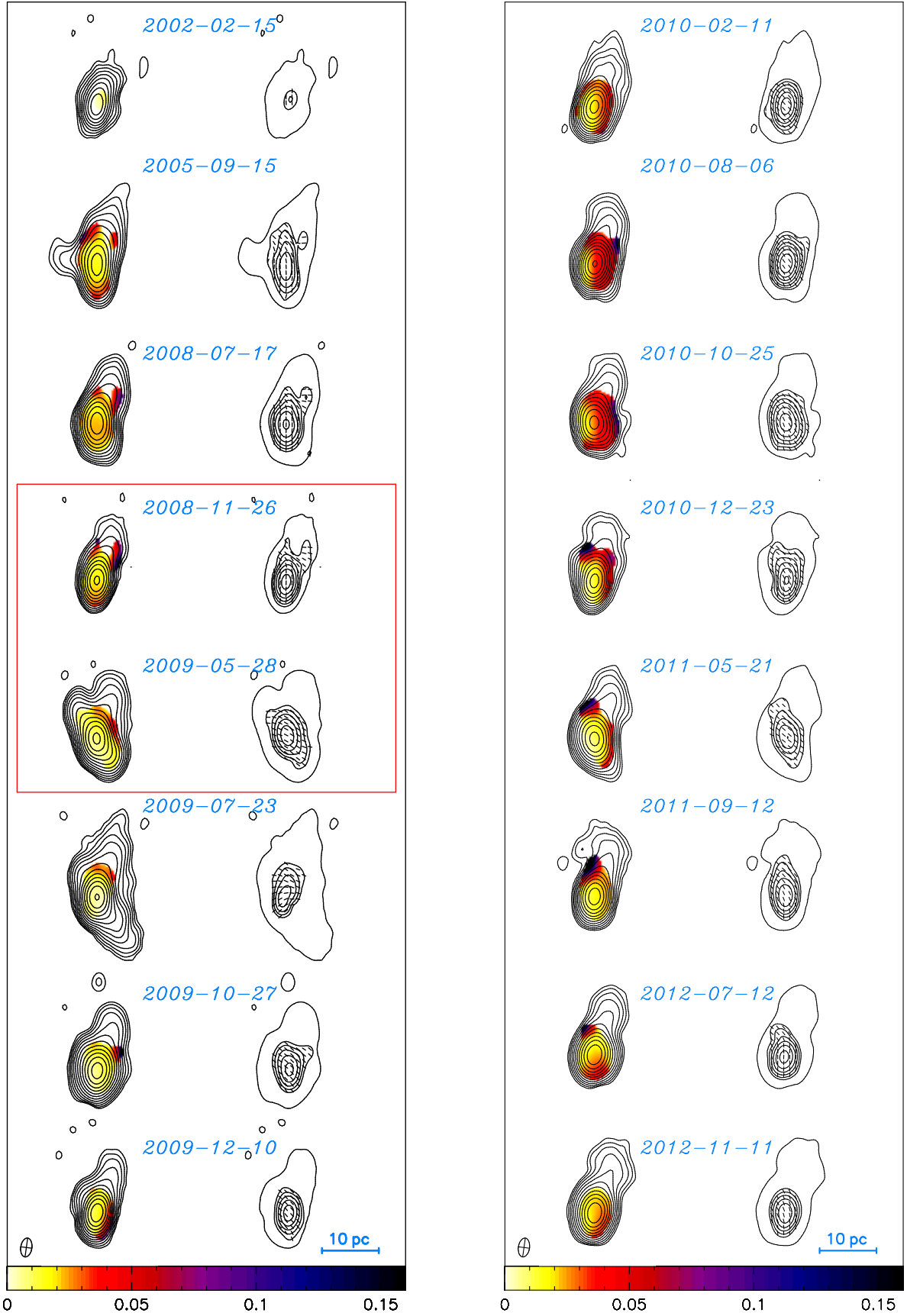

 \centering
 \resizebox{0.38\hsize}{!}{\includegraphics[angle=-90,width=0.36\textwidth]{figs/2023_panel_left.ps}}\hspace{1.5cm}
 \resizebox{0.38\hsize}{!}{\includegraphics[angle=-90,width=0.36\textwidth]{figs/2023_panel_right.ps}}
 \caption{Sequence of 15.4~GHz VLBA images showing structural changes
   in the quasar 2023+335 during a 10.8~yr period.  The first two
   images are non-MOJAVE image from the VLBA archive.  The most
   dramatic parsec-scale variations occurred between 2008.9 and
   2009.4 marked by a red box. For each epoch we present two images. The first
   image is a naturally weighted total intensity map with linear
   fractional polarization overlaid according to the color wedge. The
   second image includes the lowest positive total intensity contour
   from the first image, and linearly polarized intensity contours
   together with electric polarization vector directions. The crossed
   ellipse in the lower left represents the FWHM of the restoring beam of
   $0.96\times0.56$~mas at $\mathrm{PA} = -8\fdg2$, which is the
   median for all restoring beams at the 16 epochs. One milliarcsecond
   corresponds to about 3.5~pc at the source redshift of $z = 0.22$.
   The image parameters are listed in Table~\ref{t:maps}. 
   }
 \label{f:maps}
\end{figure*}

\begin{figure*}
\centering
 \resizebox{0.85\hsize}{!}{\includegraphics[angle=-90]{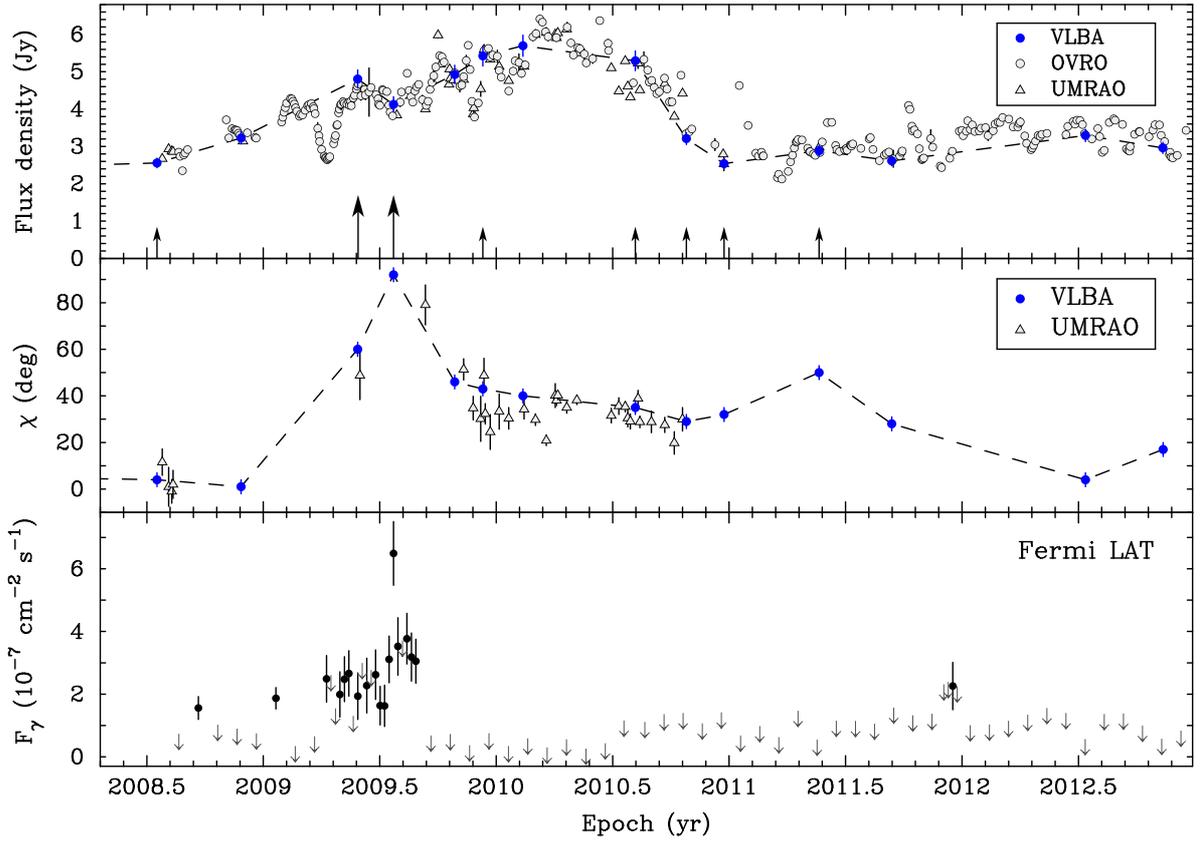}}
 \caption{Top: 15.4~GHz total flux density VLBA (filled circles), 15~GHz OVRO (unfilled circles), 
          and 14.5~GHz UMRAO (unfilled triangles) light curves. The epochs of VLBA 
          observations with signatures of multiple imaging (see Fig.~\ref{f:maps}) are marked 
          by large (strong manifestation) and small arrows (weaker effect)
          with distinction made visually from Fig.~\ref{f:maps} on the basis of the
          extension of the refraction-induced parsec-scale core sub-structure.
          Middle: electric vector position angle evolution as measured by 15.4~GHz VLBA 
          (filled circles) and single-dish 14.5~GHz UMRAO (unfilled triangles) observations. 
          Bottom: integrated $0.1-200$~GeV photon flux measured with the {\it Fermi} LAT.
          Downward arrows indicate $2\sigma$ upper limits.
         }
 \label{f:light_curves}
\end{figure*}

\begin{figure*}
 \centering
 \resizebox{0.9\hsize}{!}{\includegraphics[angle=-90]{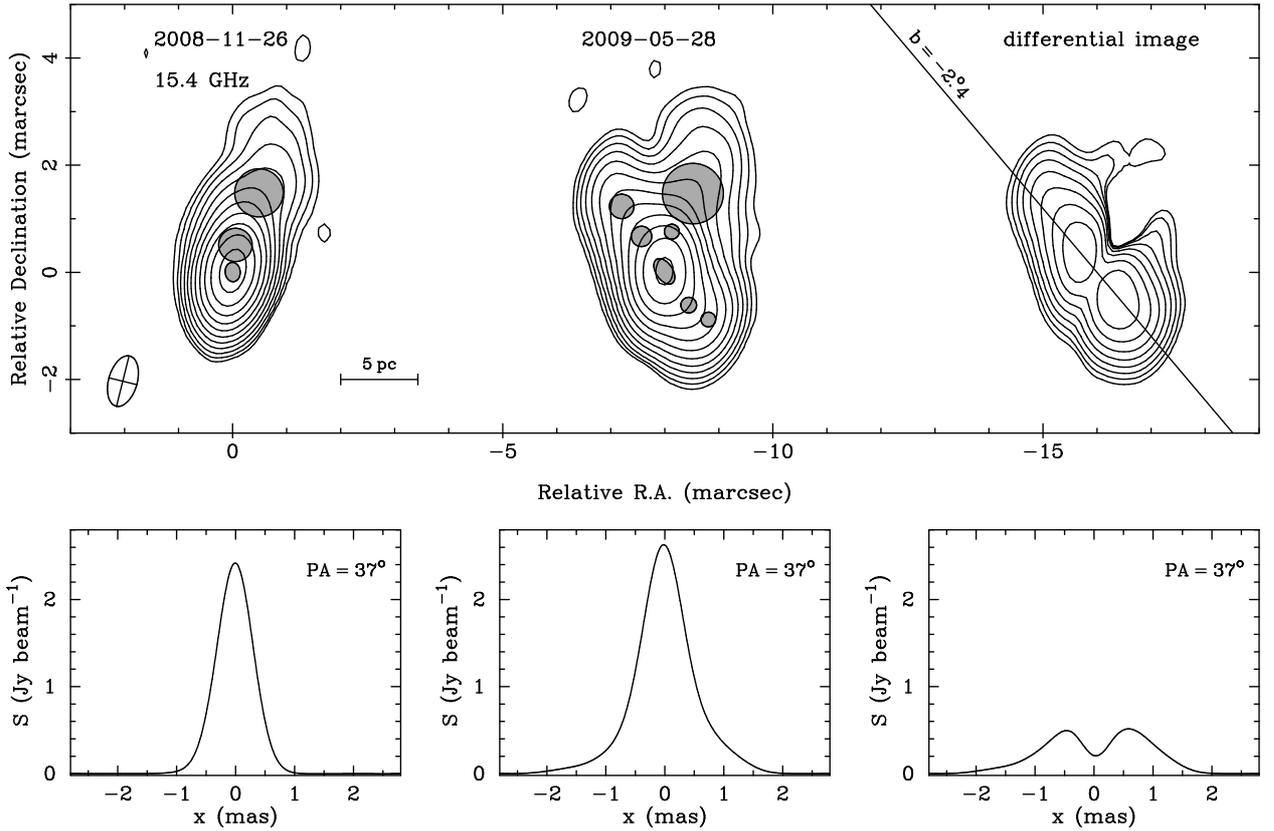}}
 \caption{Extraction of the refraction-induced structure from the
   naturally weighted 15.4~GHz VLBA images of 2023+335.  The image on
   2008 November 26 (top left) with the peak of 2.42~Jy~beam$^{-1}$
   shows a typical parsec-scale morphology represented by a bright
   core and one-sided jet propagating along the
   $\mathrm{PA}\sim-20\degr$.  The image taken on 2009 May 28 (top
   middle) with a peak flux density of 2.62~Jy~beam$^{-1}$ shows an
   unusual brightness distribution, which is a result of multiple
   imaging of the source. The fitted Gaussian components are
   superposed on the images as shaded circles/ellipses. The difference 
   image of the two epochs (top right) reveals a quasi-symmetric pattern
   dominated by two bright components and formed by refraction when
   the lens edge passes over the source. The pattern of de-magnified
   secondary images is extended nearly precisely along the line of
   constant galactic latitude, which is at $\mathrm{PA}\approx40\degr$
   in this region of the sky (full line).  All the images were
   convolved with an identical restoring beam, the FWHM of which is
   depicted as the crossed ellipse in the lower left corner. One
   milliarcsecond corresponds to about 3.5~pc. The total intensity
   profiles (bottom panel) are shown along the $\mathrm{PA}=37\degr$ 
   that connects the peaks of the refractive pattern in the difference 
   image. The angular separation of the peaks is about 1~mas.  }
 \label{f:diff_image_u}
\end{figure*}

\begin{figure*}
 \centering
 \resizebox{0.9\hsize}{!}{\includegraphics[angle=-90]{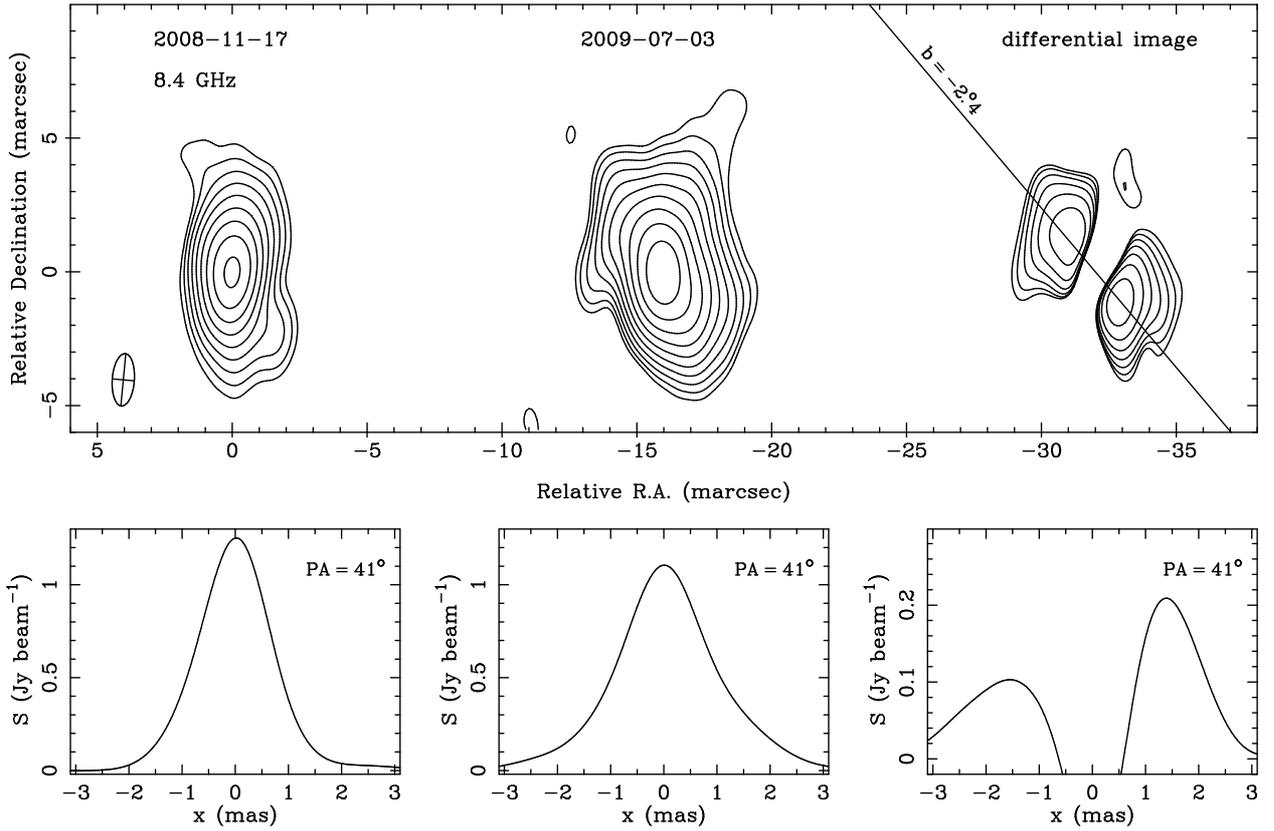}}
 \caption{Same analysis as in Fig.~\ref{f:diff_image_u} but for 
          different epochs and at 8.4~GHz.
         }
 \label{f:diff_image_x}
\end{figure*}

\subsection{Serendipitous discovery of unusual parsec-scale structure of the quasar at 15~GHz}

In Fig.~\ref{f:maps}, we present naturally weighted total intensity
and linear polarization VLBA images of 2023+335 at 16 epochs covering
a time period of about 11 years. Since the source was not observed in
the MOJAVE program prior to 2008, we have reduced the available
archival VLBA data at 15~GHz to examine the past structural history of
the source.  The image parameters are listed in Table~\ref{t:maps}.

Typically, the milliarcsecond-scale radio morphology of bright, compact 
AGN jets consists of a one-sided core-jet structure,
reflecting strong selection effects and the Doppler boosting of jet
emission \citep[e.g.,][]{MOJAVE_VI}.  The VLBI core component is
associated with the apparent origin of AGN jets and commonly appears
as the brightest feature in VLBI images. The core is typically
unresolved along one or both axes of the restoring beam
\citep{Kovalev05}.

\begin{table}
\caption{Parameters of the 15 GHz naturally weighted images.}
\label{t:maps}
{\centering
\renewcommand{\footnoterule}{}
\begin{tabular}{c c c c c r}
\hline\hline
Epoch        & $S_\mathrm{VLBA}$ & $S_\mathrm{base}$ & $P_\mathrm{VLBA}$ & $P_\mathrm{base}$ & $\chi$     \\
             & (mJy)             & (mJy bm$^{-1}$)   & (mJy)             & (mJy bm$^{-1}$)   & ($^\circ$) \\
(1)          & (2)               & (3)               & (4)               & (5)               & (6)        \\ 
\hline
2002--02--15 & 1715 & 2.5 &   6 & 2.6 &$-$4 \\
2005--09--15 & 2041 & 1.2 &  28 & 1.0 &   9 \\
2008--07--17 & 2557 & 0.8 &  52 & 0.8 &   4 \\
2008--11--26 & 3228 & 1.5 &  53 & 0.7 &   1 \\
2009--05--28 & 4806 & 1.1 &  48 & 0.9 &  60 \\
2009--07--23 & 4125 & 0.6 &  16 & 0.9 &  91 \\
2009--10--27 & 4952 & 1.0 &  42 & 1.2 &  46 \\
2009--12--10 & 5442 & 2.3 &  86 & 2.5 &  43 \\
2010--02--11 & 5696 & 1.5 & 120 & 1.2 &  40 \\
2010--08--06 & 5294 & 0.9 & 183 & 1.2 &  35 \\
2010--10--25 & 3213 & 1.4 &  99 & 1.1 &  29 \\
2010--12--24 & 2539 & 1.3 &  44 & 0.8 &  32 \\
2011--05--21 & 2884 & 1.2 &  40 & 1.1 &  50 \\
2011--09--12 & 2616 & 1.0 &  44 & 0.9 &  28 \\
2012--07--12 & 3294 & 1.1 &  68 & 1.1 &   4 \\
2012--11--11 & 2960 & 1.0 &  65 & 1.3 &  17 \\
\hline
\end{tabular}
}

{\bf Notes.} Columns are as follows:
(1) epoch of observations,
(2) total VLBA flux density,
(3) lowest contour in total intensity image,
(4) total VLBA polarized intensity,
(5) lowest contour in polarization intensity image,
(6) position angle of the electric vector on the sky.
\end{table}

When 2023+335 was first added to the MOJAVE VLBA imaging program in
2008, the source displayed a classical one-sided morphology, with a
strong core and a jet extending along position angle $\sim-20\degr$.
However, an image taken in May 28, 2009 unexpectedly showed surprising
structural changes, with additional bright new emission regions
elongated in a direction of position angle $\sim40\degr$ and
$\sim-130\degr$. We have not witnessed this type of behavior in any of
the other $\sim300$ AGN jets regularly monitored by the MOJAVE or VLBA
2cm survey programs \citep{2cmPaperI,2cmPaperII,MOJAVEI,MOJAVE}. Such
unusual structural evolution cannot be explained by the standard
relativistic jet expansion model \citep{Blandford79} since it would
require bi-directional superluminal expansion ~--- an impossibility
according to simple causality arguments. Thus, we rule out the
intrinsic variation scenario.

\subsection{Galactic plane sky location}

The first clue to the physical mechanism responsible for these
dramatic and quite atypical morphological changes is the fact that
the source is at very low galactic latitude $b=-2\fdg37$, suggesting
that those structural changes could be caused by some form of
propagation effect. Light propagation effects have been seen in other
AGN, in the form of scintillation \citep[see
e.g.,][]{Quirrenbach89,Jauncey00,Dennett-Thorpe02,Lovell08,Savolainen08}
or extreme scattering events \citep{Fiedler87,Cimo02,Senkbeil08}. The 
second hint is the galactic longitude $l=73\fdg13$ of the source, which
places it behind the highly turbulent Cygnus region
\citep{Bochkarev85,Fey89}. The line of sight to 2023+335 passes near
the \object{Cygnus} loop supernova remnant \citep{Pineault90} that may locally
ionize the ISM \citep{Romani87}. 

\subsection{Extreme scattering events}

The final clue to the event can be found in the OVRO light curve
(Fig.~\ref{f:light_curves}, top panel). This shows that the source was
undergoing an ESE around the epoch 2009.3 due to the prominent dip and 
rise features, as was previously reported by \cite{Kara12}. We can 
additionally conclude that the refractive
strength of the scattering screen must be very high, because all
previously known ESEs have been detected at lower radio frequencies,
at which the effect is more pronounced. The VLBA observations on May
28, 2009 serendipitously coincided with a special feature of the ESE,
the caustic spike, when the secondary images are expected to have the
largest angular separations
\citep{Fiedler87,Romani87,Clegg98,Lazio04}. All these facts coherently
lead to the conclusion that the new-born structure detected on May 28,
2009, is the result of multiple imaging of the source, induced by
refraction in an intervening ISM screen. We model the physical
characteristics of the scattering screen in Sect.~\ref{s:ESE}.

In the current paradigm, an ESE occurs when a localized
structure (lens/cloud) of partly-ionized material with electron
density enhancements passes in front of a distant background radio
source. The significant changes in flux density attributed to ESE
are expected when the ratio of the angular size of the lens to the
intrinsic angular size of the VLBI core is within a quite narrow
range $[1/n; n]$, where $n$ is limited by a few. This condition, as
well as the turbulence spectrum of a typical screen, determines the
expected rate of such events.  In addition, the electron density in
the screen must be high enough to result in significant scattering
that can be detected and recognized as an ESE.  ESEs are more
prominent at lower frequencies (a few GHz), as the refractive power
of the lens is greater. This is why most ESEs are detected at 2~GHz,
some at 6--8~GHz, and only one, to our knowledge, at 15~GHz (this
paper).

In the case of 2023+335, which is a galactic plane source seen through
the Cygnus region, the probability of observing an ESE is certainly
higher than average. For the successful detection of more frequent 
ESEs, the source should be monitored at lower frequency. Yet even at 
15~GHz, the OVRO monitoring revealed a clear ESE after only one year 
of the operation of the program, and has provided evidence for other 
ESEs in the form of flux density excursions at later epochs (around 
$\sim2009.9$, $\sim2010.05$, $\sim2011.2$, $\sim2011.7$), indicating 
more complex structure of the scattering screen. Thus, the quasar 
2023+335 is a promising target for future ESE detections and 
corresponding studies.

\subsection{Joint analysis of VLBA, OVRO, and UMRAO 15~GHz data 
supplemented by $\gamma$-ray Fermi results}

In order to construct an overall picture of what has occurred in the
source since $\sim$2008.5, we analyzed our 15.4~GHz total VLBA flux
density measurements, superposed with the more densely sampled OVRO
and UMRAO light curves (Fig.~\ref{f:light_curves}, top panel). We
found good agreement between (i) the VLBA and single-dish flux density
measurements, and (ii) the VLBA and UMRAO EVPAs, indicating that
virtually all of the emission of 2023+335 at 15~GHz originates from
parsec scales probed by the VLBA.  The source was undergoing a flare
during the period 2008.5--2010.5, during which it increased its
intensity by a factor of $\sim$2. Most likely, the flare was intrinsic
to the source, rather than the result of a focusing effect caused by
the ISM. The radio flare is expected to occur when the
perturbation moving down the jet crosses the $\tau_\nu\sim1$ region
at a given frequency $\nu$, the VLBI core.  This happens during
2010, when the flux density reaches the maximum of about 6~Jy.
Therefore, in 2009 the perturbation is still within the compact VLBI
core region, not at significantly larger angular separations, where
the new-born emission is detected in May 2009. Moreover, the
perturbation is expected to propagate in the direction of the jet
($\mathrm{PA}\sim-20\degr$), while the observed morphology instead
shows the new-born emission appearing at $\mathrm{PA}\sim40\degr$
and $\mathrm{PA}\sim-130\degr$.  

The intrinsic core flare scenario is supported by several
observations.  First, the flare was accompanied by a rapid
$\sim90\degr$ change in EVPA (Fig.~\ref{f:light_curves}, middle
panel). This is typically observed when a perturbation crossing the
$\tau_\nu=1$ core region at a given frequency, i.e., the VLBI core,
changes the regime of synchrotron radiation from being optically thick
to optically thin, and the EVPA subsequently returns slowly to the
pre-flare value, $\chi\sim0\degr$, that roughly aligns with the inner
jet direction.  Second, it is known that the $\gamma$-ray emission is
not affected by the ISM, but during the rising part of the radio
flare, the source became bright at energies above 100~MeV
(Fig.~\ref{f:light_curves}, bottom panel), with the $\gamma$-ray peak
preceding the 15~GHz one by about 8 months \citep[see also][]{Kara12}.
This is consistent with the established close radio/$\gamma$-ray
connection \citep[e.g.,][]{Kovalev09,Pushkarev10} in the light of a
scenario of broadband synchrotron-Compton flares in the base of the
jet \citep[e.g.,][]{Dermer09,Bottcher10}. Therefore, the flaring in
the 15~GHz VLBA core, the apparent jet base, made this most-compact
region of the source even more flux-density dominated and thus a
region more susceptible to propagation effects in the ISM. The latter
are more pronounced for highly compact features.

To analyze the refractive-induced pattern detected on May 28, 2009 in
more detail we fit the observed brightness distribution at this epoch
with Gaussian emission features (Table~\ref{t:model}) and performed
difference imaging with the previous epoch, 2008 November 26, which
appears to be the least affected by scattering
(Fig.~\ref{f:diff_image_u}). The pattern is quasi-symmetric, and
consists of a hierarchy of secondary images produced by ray crossings.
The best fit model contains four Gaussian components that are
progressively de-magnified with angular separation from the primary
image, with two closer sub-images at the level of $\sim$10\% of the
core flux density and two outer ones at $\sim(3-4)$\% level. The
pattern is stretched out along the line of constant galactic latitude.
A noteworthy feature of the pattern is that it exhibits sub-images on
either side of the core, while in a typical ESE event these are
expected to lie only on one side \citep{Clegg98}. This could be 
explained by complex structure of the screen, having free-electron 
density enhancements separated by an angular distance comparable to 
the VLBI core extent, i.e. a double-lens system, like those investigated 
by \cite{Kim05}. In this case, two secondary images could be created on 
both sides of the VLBI core, if the density profile of each lens is close 
to Gaussian \citep{Clegg98}.  The primary image would likely show some 
elongation. This is consistent with the observed ellipticity of the VLBI 
core component (Fig.~\ref{f:diff_image_u}, Table~\ref{t:model}).

\begin{table}
\caption{15 GHz Gaussian component model at epoch 2009 May 28.}
\label{t:model}
{\centering
\renewcommand{\footnoterule}{}
\begin{tabular}{crrcrccr}
\hline
\hline
Comp.&  $S$ & SNR &  $r$ & $\varphi$ & Maj. & Ratio & P.A. \\
     &(mJy) &     &(mas) &     (deg) &(mas) &       &(deg) \\
 (1) &  (2) & (3) &  (4) &       (5) & (6)  &   (7) &  (8) \\
\hline
  C  & 3649 & 833 &\ldots&    \ldots & 0.53 &  0.58 &    33 \\
  J  &   51 &  18 & 0.75 &     $-$11 & 0.28 &  1.00 &\ldots \\
  J  &  121 &  17 & 1.54 &     $-$20 & 1.13 &  1.00 &\ldots \\
  S  &  393 & 166 & 0.77 &    $-$144 & 0.29 &  1.00 &\ldots \\
  S  &  302 &  70 & 0.78 &        33 & 0.38 &  1.00 &\ldots \\
  S  &  144 &  83 & 1.21 &    $-$138 & 0.27 &  1.00 &\ldots \\
  S  &  119 &  28 & 1.45 &        33 & 0.45 &  1.00 &\ldots \\
\hline
\end{tabular}
}

{\bf Notes.} Columns are as follows:
(1) component designation: C is the core, J is the jet, S is the secondary image component,
(2) fitted Gaussian flux density at 15 GHz,
(3) signal-to-noise ratio determined as the ratio of the peak flux density of the component to the rms noise under it,
(4) position offset from the core component,
(5) position angle of the component with respect to the core component,
(6) FWHM major axis of the fitted Gaussian,
(7) axial ratio of the fitted Gaussian,
(8) major axis position angle of the fitted Gaussian.
\end{table}

We did not detect any significant image wander of the type expected
for refractive ESE \citep{Clegg98}. Our observations were not carried
out in a phase-reference mode, and during imaging we performed
self-calibration that erases all absolute position information, making
relative position shift analysis impossible. On the other hand, as the
lens covered only the most compact component, the VLBA core, it did
not affect the position of the rest of the real jet structure, which
we were able to use as the reference location when applying the
two-dimensional cross correlation technique \citep{Lewis95}.

The signatures of refractive sub-images at epochs 2008 July 17, 2009
May 28, 2009 July 23, 2009 December 10, 2010 August 6, 2010 December
23, and 2011 May 21 are all preferentially aligned along the constant
galactic latitude line with respect to the VLBI core
(Fig.~\ref{f:maps}). This indicates that the discrete structures of
electron density enhancements acting as refractive lenses move
parallel to the Galactic plane, which corresponds to the direction
from North-East to South-West on the images in the equatorial
coordinates. This is consistent with the changes in the induced
patterns between the epochs of May and July 2009, implying that the 
observed proper motion of the screen is prograde with regard to the 
galactic center. However, emission with a high signal-to-noise ratio 
was detected due East of the core at one full-track pre-MOJAVE epoch 
with high sensitivity, September 15, 2005, presumably suggesting that 
a lens drifted over the source along the tangent line.  The remaining 
epochs, with the possible exception of November 26, 2008, are 
characterized by angular broadening of the VLBI core region.  
Interestingly, the secondary images that are also detected in linear 
polarization, e.g., on May 28, 2009, December 23, 2010, and May 21, 
2011, show EVPAs aligned with those in the primary image.%, as expected.

\subsection{Additional evidence from 8 GHz VLBA observations}

The quasar 2023+335 was also observed with the VLBA at 8.4~GHz on five
occasions from November 2008 to November 2009 as a phase-reference
calibrator in a framework of the black hole X-ray binary V404 Cyg
project \citep{Miller-Jones09}. We reduced these archival VLBA data
and found structural behavior very similar to our 15.4~GHz results.
The source was least affected by the ISM in November 2008, but
revealed multiply-imaged structure stretched out along the constant
galactic latitude line at other epochs, including the 3 July 2009
epoch (Fig.~\ref{f:diff_image_x}), which was only 20 days prior to our
15.4~GHz MOJAVE epoch. The 8.4~GHz peaks of the induced structure are
separated by $\sim$3.0~mas, while at 15.4~GHz the separation is
$\sim$0.8~mas on 23 July 2009 ($\sim$1.0~mas on 28 May 2009). The
separation thus scales as $\lambda^{2.2}$ ($\lambda^{1.75}$ 28 May 2009), 
which we consider to be convincing proof of a plasma scattering origin 
for the multiple imaging event \citep{Rickett88}, because the refraction 
angle in plasma scattering is expected to follow a $\lambda^2$ dependence 
\cite{Clegg98}, while for the intrinsic scenario the studied parameter 
of the angular separation is frequency-independent since the synchrotron
jet emission is optically thin \cite[e.g.,][]{PK2012}.

\subsection{Summary of observational evidence in support of refractive scattering}

To summarize, we have found several pieces of observational
evidence which suggest that the multiple imaging event witnessed in
2009 in 2023+335 was caused by refraction in the interstellar
medium:
\begin{itemize}

\item the 8 and 15~GHz parsec-scale jet structure shows strongly
  induced patterns at two epochs, with weaker manifestations of the
  effect seen at several other epochs;

\item the strongly induced patterns are coincident with an ESE event
  in the radio light curve, as predicted by models;

\item the source is located in the Galactic plane behind the Cygnus
  loop, which provides a high probability of a propagation effect
through an intervening turbulent screen;

\item the angular separation between the peaks of the induced structure 
  shows a $\lambda^2$-dependence, which is convincing proof of a plasma 
  scattering origin of the induced sub-images. This argument also rules 
  out a scenario in which the observed structural changes between epoch 
  of Nov 2008 and May 2009 are intrinsic to the source.
\end{itemize}

\section{Derived properties of the screen}
\label{s:ESE}

\begin{figure}
 \resizebox{\hsize}{!}{\includegraphics[angle=-90]{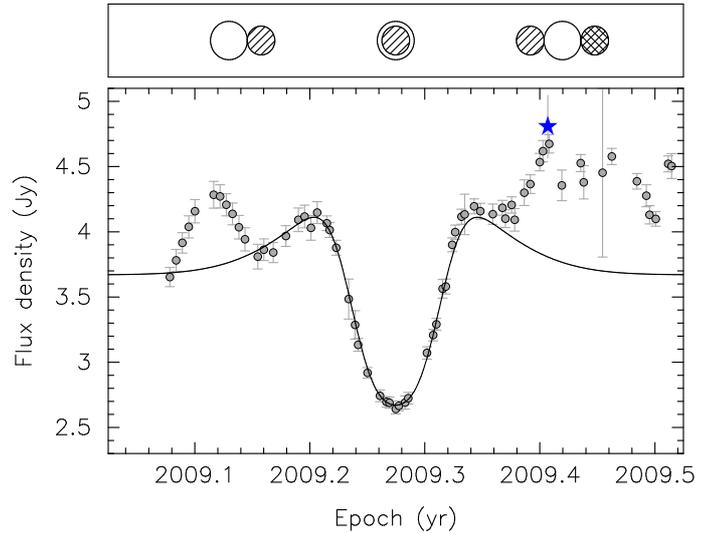}}
 \caption{Bottom: 15 GHz OVRO light curve in the quasar 2023+335 during an extreme 
          scattering event. The solid line shows the stochastic broadening 
          model light curve for the ESE. A two-component model was used: a 
          lensed component of flux density 1.87~Jy, and an un-lensed component 
          of flux density 1.80~Jy. The star symbol represents the MOJAVE epoch 
          that revealed multi-component structure induced by refraction (see 
          Fig.~\ref{f:diff_image_u}). This epoch of VLBA observations 
          serendipitously coincided with the local sharp spike, possibly 
          ascribed to the outer caustic that is associated with the passage of 
          the lens edge over the background source. Top: a toy model of the 
          scattering screen during the ESE. The unfilled circle represents the 15.4~GHz 
          VLBA core of the background quasar. The hatched circle represents the lens that 
          passes over the source. The cross-hatched circle represents another electron
          density enhancement of the screen.}
 \label{f:ESE_modeling}
\end{figure}

We can draw some basic conclusions from a simple analysis of the shape
of the ESE light curve (Fig.~\ref{f:ESE_modeling}). The rounded
minimum of the ESE suggests that the lens is comparable to or smaller in
size than the part of the source which it occults; otherwise, a
flat-bottomed minimum is expected. Also, the ESE caused a $\sim30$\%
decrease in the source's total flux density at 15~GHz.  This indicates
that the lensed part of the source contains a significant fraction of
the total flux density. The most likely candidates for such a region
are the VLBI core and/or, if present and bright, the innermost VLBI
jet component, because (i) the source is highly core dominated on
milliarcsecond scales and (ii) the total VLBA flux density agrees well
with the single-dish flux density of the source, indicating that
virtually all the 15~GHz emission originates from milliarcsecond
scales. Finally, the sharp spike at the epoch of $\sim$2009.4 is
likely attributable to an outer caustic that is associated with the
passage of the lens edge over the background source. The presence of
this caustic suggests that the refractive scattering is strong, and
multiple images are expected to be formed (Fig.~\ref{f:diff_image_u}).
The other outer caustic at the epoch of $\sim$2009.1 is present but
less pronounced, indicating a difference in the free-electron density  
profile across the scattering screen.

\subsection{Stochastic broadening}

To derive the quantitative parameters of the plasma lens, we used a
statistical model for flux redistribution developed by
\cite{Fiedler94} and based on stochastic broadening regardless of its
nature, refractive or diffractive. In this model, the flux density of
a distant background source at a time $t$ during an ESE is determined
as $f(t; I_0,\mu,\theta_s,\theta_l,\theta_b)$, where $I_0$ is the
nominal (un-lensed) flux density level of the source outside the lens,
$\mu$ is the proper motion of the lens across the line of sight,
$\theta_s$ is the intrinsic FWHM angular size of the source,
$\theta_l$ is the apparent angular width of the lens (which has a
band-shaped geometry when projected on the sky), and all parallel rays 
incident on the lens are assumed to be scatter-broadened into a Gaussian 
brightness distribution of FWHM $\theta_b$. Both $\theta_l$ and $\theta_b$ 
are measured with respect to $\theta_s$; thus we fit their ratios.

We used a two-component model, with a lensed component
$I_\mathrm{scat}(t)$ being subject to temporal flux density variations
from scattering during the ESE, and an un-lensed part
$I_\mathrm{unscat}$ such that their sum equals the nominal flux
density level outside the ESE.  We also assumed that there are no
significant intrinsic flux density changes during the event. We then 
fitted $\theta_l$, $\theta_b$, $I_0$, $I_\mathrm{scat}$, and $\mu$ by
searching in parameter space and evaluating the goodness of fit using the
$\chi^2$ statistic. We found good agreement with the depth and shape
of the minimum being reproduced.  In Fig.~\ref{f:ESE_modeling}, we
show the simulated light curve superposed on the 15~GHz OVRO light
curve for the set of best-fit parameters listed in
Table~\ref{t:ESE_modeling}. The scattered component comprises 51\% of
the total flux density.  This rules out a scenario where the lens
drifts over any downstream jet component, because the brightest of
them, at the latest prior epoch (2008 November 26), accounted for only
about 10\% of the 15~GHz total flux density. Therefore, it is most
likely that the lens passed over the VLBI core component.

\begin{figure}
 \resizebox{\hsize}{!}{\includegraphics[angle=0]{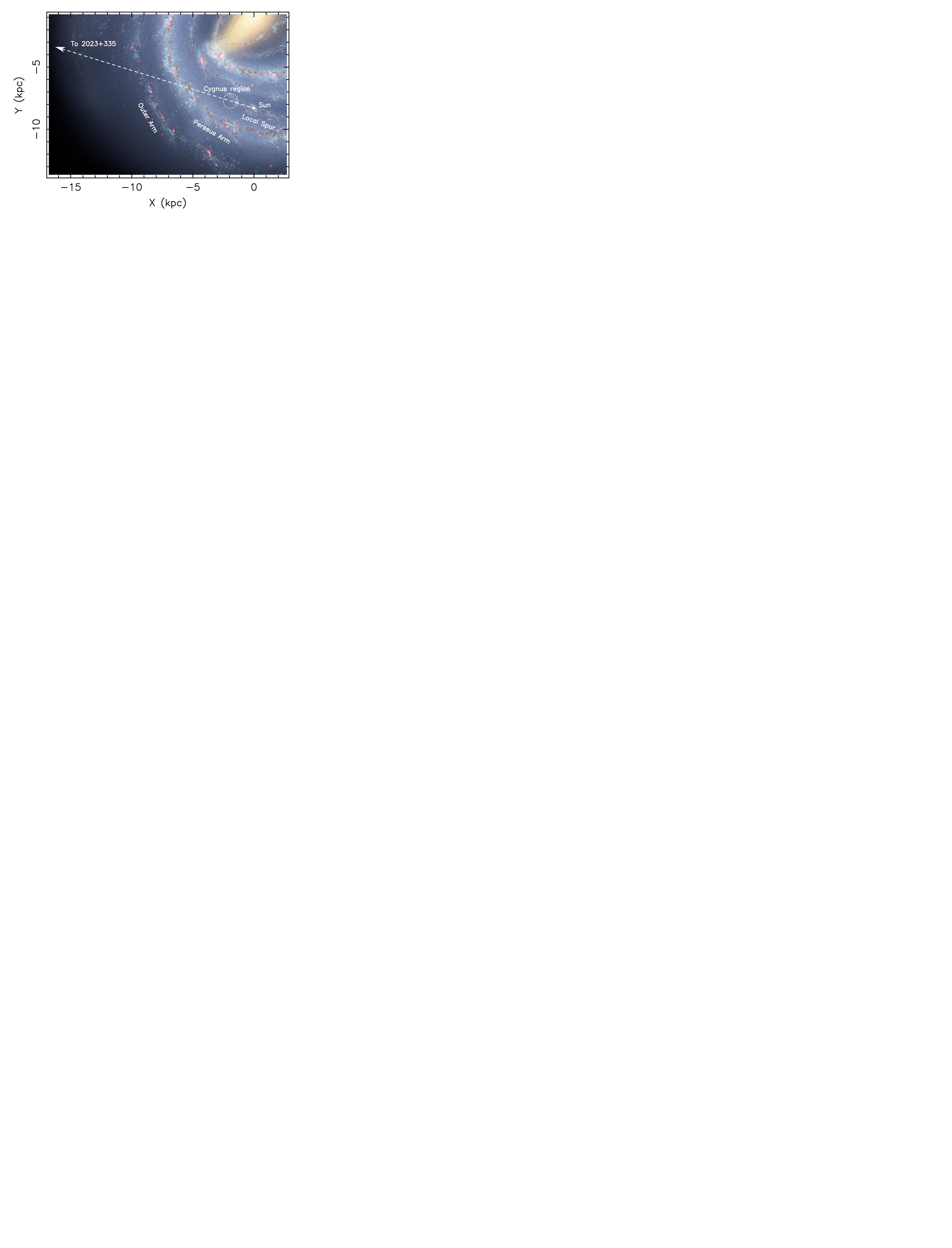}}
 \caption{Artists' impression of a face-on view of the Galactic plane
   (image credit: R. Hurt, NASA/JPL-Caltech/SSC).  The Sun is placed
   at 8.3~kpc \citep{Gillessen09} from the Galactic center $(0,0)$.
   The dashed line shows a direction toward 2023+335 with a galactic
   longitude $l=73\fdg13$ that passes through the nearby
   ($D\sim1.5$~kpc) Cygnus region in the Local Arm, the Perseus Arm
   ($D\sim6$~kpc), and the Outer Arm ($D\sim10$~kpc). We assume that
   the lens is located in the turbulent Cygnus region (dotted circle)
   at the distance of $\sim1.5$~kpc.  }
      \label{f:gal_arms}
\end{figure}

The fitted parameters could be used to calculate the following 
physical characteristics of the plasma lens:
\begin{itemize}
\item Angular size $\theta_l=\theta_s/1.03$, where the intrinsic
  angular size of the VLBI core
  $\theta_s=\theta_b/1.88\approx0.28$~mas, and
  $\theta_b\approx0.53$~mas is the FWHM of the observed VLBI core size
  at epoch 2009 May 28. We then obtained $\theta_l\approx0.27$~mas,
  which is an upper limit since the lens passage would most likely
  occur along a chord rather than precisely along the diameter of the
  VLBI core.  It is also worth noting that the observed size of the
  VLBI core component at the epoch 2008 November 26, which seems to
  show the source structure when least affected by scattering, is only
  $\sim$15\% broader than the inferred intrinsic extent
  $\theta_s=0.28$~mas.
\item Transverse linear size $a=\theta_l\,D$, where $D$ is the
  distance to the lens. The distance to the lens is uncertain, as the
  line of sight to 2023+335 passes through the nearby ($D\sim1.5$~kpc,
  e.g., \citealt{Rygl2012}) Cygnus region in the Local Arm (also known as 
  the Orion or Local Spur), the Perseus Arm ($D\sim6$~kpc), and the
  Outer Arm ($D\sim10$~kpc) (see Fig. \ref{f:gal_arms}).  We assume
  that the lens is in the highly turbulent Cygnus region at the
  distance of $\sim1.5$~kpc, which yields $a\approx0.4$~AU.
\item Proper motion $\mu=3.5\,\theta_s/\tau\approx6.8$~mas~yr$^{-1}$, 
  where $\tau\approx0.14$~yr is the duration of the ESE as measured 
  between the peaks that surrounds the minimum of the flux density 
  modulation. 
\item Transverse speed $V_{l\perp}=4.74\,\mu_{\mathrm{[mas/yr]}}\,D_{\mathrm{[kpc]}}\approx48.7$~km~s$^{-1}$ 
  with respect to the observer. Taking into account the Earth's motion 
  around the Sun, the projected speed of the screen at the middle 
  phase of the ESE in the Sun's rest frame is about 56~km~s$^{-1}$. 
\end{itemize}

The fitted light curve reproduces the inner caustics quite well.
What the model does not reproduce are the flux density enhancements 
attributed to outer caustics, the earlier one of which shows an excess of 
$\sim$0.5~Jy comparing to the unlensed flux density level, while the later 
outer caustic is characterized by nearly twice as large flux density excess, 
$\sim$1~Jy. The latter value is consistent with the total flux density of 
the induced quasi-symmetric multi-component pattern (see 'S' components in 
Table~\ref{t:model}), indicating that the scattering screen at epoch 2009 
May 28 has a double-lens structure (Fig.~\ref{f:ESE_modeling}, upper panel) 
that creates the secondary images on both sides from the VLBI core, as 
depicted in Fig.~\ref{f:diff_image_u}, and thus considerably deviates from 
the model light curve.

\begin{table}
\caption{Parameters of the best-fit stochastic broadening model.}
\label{t:ESE_modeling}
\begin{center}
\renewcommand{\footnoterule}{}
\begin{tabular}{lcr}
\hline\hline
Parameter              & Value &                 Unit \\
\hline
$\theta_s/\theta_l$    &  1.03 &                      \\
$\theta_b/\theta_s$    &  1.88 &                      \\
$I_0$                  &  3.67 &                   Jy \\
$I_\mathrm{scat}/I_0$  &  0.51 &                      \\
$\tau^\mathrm{\,\,a}$  &  0.14 &                   yr \\
$\mu$                  & 25.01 & $\theta_s$ yr$^{-1}$ \\
$\chi^2$               &  0.02 &                      \\
\hline
\end{tabular}
\end{center}
{\bf Note.} $^\mathrm{a}$ This parameter held fixed.
\end{table}

\subsection{Refraction defocusing}

Another approach is provided by the refraction defocusing (RD) model
\citep{Clegg98}, in which the intervening scattering screen is
represented by a plasma lens with a Gaussian profile of free-electron
column density. The refractive properties of the lens in this model
are described completely by an introduced dimensionless parameter
\begin{equation}
\alpha=3.6\left(\frac{\lambda}{\mathrm{1~cm}}\right)^2
          \left(\frac{N_0}{\mathrm{1~pc~cm}^{-3}}\right)
          \left(\frac{D}{\mathrm{1~kpc}}\right)
          \left(\frac{a}{\mathrm{1~AU}}\right)^{-2}\,,
\end{equation}
where $N_0$ is the maximum column density through the lens and
$\lambda$ is the observing wavelength. In case of moderate and strong
refraction, the caustic spikes are expected to be formed. The relative
separation between the inner $\Delta x_i^\prime$ and outer $\Delta
x_o^\prime$ pair of caustics is an important observable that allows
us to place a constraint on $\alpha$ through the following analytic
expression \citep[][Eq.~23]{Clegg98}
\begin{equation}
\alpha \simeq 1.7\left(3\frac{\Delta x_o^\prime}{\Delta x_i^\prime}-1\right)^{1.06}\,.
\end{equation}
For the ESE in 2023+335, the ratio $\Delta x_o^\prime/\Delta x_i^\prime\approx2$. 
Then the strength of the lens is $\alpha\simeq10$. We can now derive the 
following properties of the lens:
\begin{itemize}
\item maximum electron column density through the lens 
      $N_0=0.28\,\alpha\,\lambda_\mathrm{[cm]}^{-2}\,D_\mathrm{[kpc]}\,\theta_{l\,\mathrm{[mas]}}^2\approx2.5\times10^{17}$~cm$^{-2}$, 
      where $\lambda$ is the wavelength of observation in cm;
\item free-electron density within the lens $n_e\approx N_0/a\approx4.0\times10^4$~cm$^{-3}$;
\item mass of the lens $M_l\sim m_p n_e a^3$, where $m_p$ is the proton mass. For a lens at 1.5~kpc, 
      $M_l\sim1.4\times10^{19}\mbox{~g}\sim7.2\times10^{-15} M_\mathrm{\sun}$.
\end{itemize}

It is worth noting that the inner caustics are less pronounced in
comparison with the outer one clearly seen at the epoch of
$\sim$2009.4 in Fig.~\ref{f:ESE_modeling}, while in the RD model they
are expected to have comparable levels. However, simulations performed
by \cite{Stinebring07} considering a two-dimensional (rather than a
one-dimensional as in the RD model) lens with a Gaussian density
profile have shown the absence of strong inner caustics. More complex
cases of double-lens systems have been investigated by \cite{Kim05}.

\section{Turbulent angular broadening}
\label{s:broadening}

Diffraction phenomena associated with scattering by a turbulent
interstellar medium results in the angular broadening of a distant
background radio source. Thus, its observed brightness distribution is
a convolution of the intrinsic source structure with a scattering
function. The angular size of a broadened source scales with frequency
through the following relation \citep{Rickett77}
\begin{equation}
\theta_\mathrm{scat}\propto\nu^{k}, \qquad k=-\frac{\beta}{\beta-2}\,,
\label{eq:scat}
\end{equation}
where $\beta$ is the index of the assumed power-law spectrum of the 
turbulent electron density fluctuations $\delta n_e/n_e$ \citep{Cordes85}
\begin{equation}
P_{\delta n_e}(q)=C_n^2q^{-\beta}, \qquad q_0\le q\le q_1\,,
\end{equation}
where $C_n^2$ is a normalizing constant, and $q_0$ and $q_1$ are
spatial wave-number cutoffs corresponding to the outer and inner
scales of the turbulence, respectively, bracketing over 6 orders of
magnitude of irregularity scales from $\sim10^{8}$~cm to
$\sim10^{15}$~cm \citep{Armstrong95,Combes00,Cordes86}. A theory of
turbulence developed by \cite{Kolmogorov41} predicts that $\beta=11/3$
for isotropic density fluctuations, and it has been reported that
observational data are consistent with this value for a number of
lines of sight \citep{Armstrong81,Wolszczan83,Rickett90,Fey91}. In
this case, the scattered angular size is expected to be proportional
to $\nu^{-2.2}$. However, the theory accommodates steeper density
fluctuation spectra ($\beta>4$) as well \citep{Cordes01}, being 
observed for some other lines of sight \citep{Hewish85,Romani86}. 
Scattering can manifest itself through diffractive and/or refractive 
effects, depending on the size of turbulent eddies and associated 
turbulence spectrum. Diffractive effects will dominate for a screen 
with small-scale electron-density irregularities, yielding shallow 
($\beta<4$) spectra, while considerably more refraction is expected 
for large-scale turbulent eddies causing steep ($\beta>4$) spectra 
\citep{Cordes86}. Note that if no scattering is present, the observed 
angular size of a flat-spectrum, inhomogeneous synchrotron source is 
expected to scale approximately as $\nu^{-1}$ \citep{KPT81}, although 
departures from this dependence are also possible and can be caused by 
pressure and density gradients in the jet or by external absorption 
from the surrounding medium 
\citep[][and references therein]{Lobanov98,Kovalev_cs,Sokolovsky_cs,Pushkarev_cs}.

\begin{table}
\caption{VLBI core parameters at different observing frequencies/epochs.}\label{t:freq_vs_size}
\begin{center}
\begin{tabular}{rccr} \hline \hline
Frequency & Epoch         & Flux density  & Angular size~ \\
(GHz)     & (yyyy-mm-dd) & (Jy)               & (mas)~~~~~~ \\
\hline
 1.421 & 1995-06-23 & $1.08\pm0.28$ & $28.49\pm7.89$ \\
 2.265 & 1995-10-12 & $1.93\pm0.44$ & $11.89\pm2.99$ \\
 2.269 & 1996-05-15 & $1.94\pm0.47$ & $11.08\pm2.99$ \\
 2.292 & 2001-03-12 & $1.20\pm0.37$ & $ 9.65\pm3.13$ \\
 2.309 & 2003-07-09 & $0.73\pm0.36$ & $14.78\pm3.75$ \\
 8.335 & 1995-10-12 & $2.66\pm0.41$ & $ 0.95\pm0.21$ \\
 8.339 & 1996-05-15 & $2.62\pm0.42$ & $ 0.74\pm0.20$ \\
 8.646 & 2001-03-12 & $1.79\pm0.34$ & $ 1.00\pm0.23$ \\
 8.646 & 2003-07-09 & $1.05\pm0.26$ & $ 1.25\pm0.37$ \\
 8.322 & 2010-01-07 & $3.17\pm0.50$ & $ 1.23\pm0.26$ \\
15.365 & 2000-04-02 & $0.89\pm0.09$ & $ 0.56\pm0.08$ \\
15.365 & 2001-09-19 & $1.54\pm0.16$ & $ 0.39\pm0.07$ \\
15.365 & 2002-02-15 & $1.27\pm0.11$ & $ 0.32\pm0.06$ \\
15.365 & 2008-07-17 & $1.27\pm0.07$ & $ 0.46\pm0.04$ \\
15.357 & 2008-11-26 & $2.80\pm0.15$ & $ 0.36\pm0.03$ \\
15.357 & 2009-05-28 & $3.65\pm0.15$ & $ 0.53\pm0.03$ \\
15.357 & 2009-07-23 & $3.44\pm0.15$ & $ 0.48\pm0.03$ \\
15.357 & 2009-10-27 & $1.87\pm0.10$ & $ 0.37\pm0.04$ \\
15.357 & 2009-12-10 & $4.87\pm0.21$ & $ 0.35\pm0.03$ \\
15.357 & 2010-02-11 & $4.52\pm0.15$ & $ 0.48\pm0.02$ \\
15.357 & 2010-08-06 & $4.30\pm0.18$ & $ 0.35\pm0.03$ \\
15.357 & 2010-10-25 & $2.05\pm0.13$ & $ 0.41\pm0.04$ \\
15.357 & 2010-12-24 & $2.16\pm0.07$ & $ 0.45\pm0.02$ \\
15.357 & 2011-05-21 & $1.08\pm0.04$ & $ 0.40\pm0.02$ \\
15.357 & 2011-09-12 & $1.32\pm0.06$ & $ 0.39\pm0.03$ \\
15.357 & 2012-07-12 & $2.51\pm0.14$ & $ 0.45\pm0.04$ \\
15.357 & 2012-11-11 & $2.31\pm0.19$ & $ 0.38\pm0.05$ \\
22.204 & 2008-06-01 & $1.39\pm0.16$ & $ 0.24\pm0.05$ \\
22.241 & 2010-04-24 & $4.71\pm0.28$ & $ 0.23\pm0.03$ \\
22.240 & 2010-07-26 & $2.97\pm0.27$ & $ 0.27\pm0.04$ \\
22.239 & 2010-09-25 & $3.03\pm0.32$ & $ 0.24\pm0.05$ \\
22.238 & 2010-11-18 & $2.41\pm0.15$ & $ 0.28\pm0.03$ \\
22.238 & 2010-12-16 & $1.37\pm0.12$ & $ 0.26\pm0.04$ \\
22.241 & 2011-04-25 & $1.64\pm0.13$ & $ 0.28\pm0.04$ \\
24.433 & 2006-06-04 & $2.05\pm0.15$ & $ 0.21\pm0.03$ \\
86.248 & 2002-04-20 & $0.42\pm0.12$ & $ 0.03\pm0.02$ \\
\hline
\end{tabular}
\end{center}
{\bf Notes.} Columns are as follows:
(1) central observing frequency, 
(2) epoch of observations, 
(3) fitted Gaussian flux density, 
(4) FWHM angular size of the fitted Gaussian.
\end{table}

\begin{figure}
 \resizebox{\hsize}{!}{\includegraphics[angle=-90]{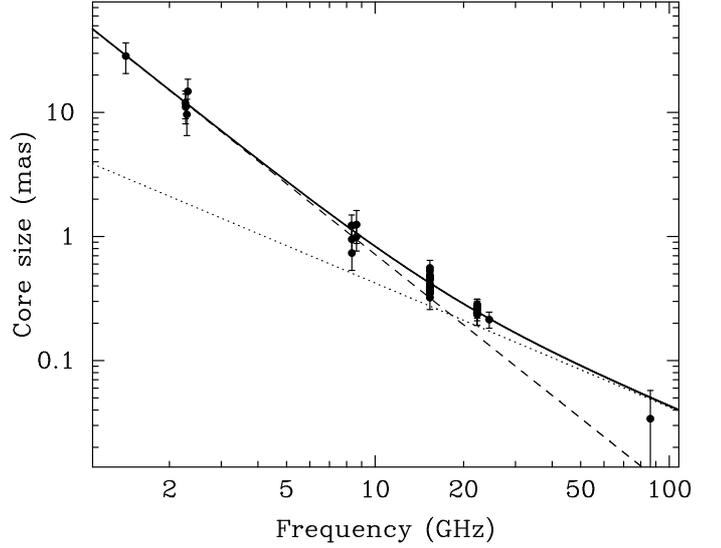}}
 \caption{Observed angular size of the VLBI core component as a function
          of observing frequency using all available VLBI data. The solid
          line represents the fit of Eq.~(\ref{r:size_vs_freq}), the dashed
          line indicates the inferred scattering diameter, and the dotted
          line shows the inferred intrinsic diameter.}
 \label{f:size_vs_freq}
\end{figure}

To investigate the amount and nature of scattering occurring along the
line of sight to 2023+335, we used the 15~GHz MOJAVE data together
with other publicly available\footnote{\url{http://astrogeo.org/vlbi_images}}
VLBI data for 2023+335 at other
frequencies, such as 1.4, 2, 8, 15, 24, and 86~GHz (see
Table~\ref{t:freq_vs_size}).  The archival data for all experiments
except two, carried out in Oct 1995 at 2 and 8~GHz \citep{Fey97} and
Apr 2002 at 86~GHz \citep{Lee08}, were reduced by us. The source
structure was model fitted with circular Gaussian components that
after being convolved with the restoring beam, adequately reproduce
the constructed brightness distribution. In Fig.~\ref{f:size_vs_freq},
we plot the angular size of the VLBI core component as a function of
observing frequency listed in Table~\ref{t:freq_vs_size}. We fitted 
the measured angular width to the following relation \citep{Lazio08}
\begin{equation}
\theta^2_\mathrm{observed} = \left(\theta_\mathrm{scat}\nu^k\right)^2 + \left(\theta_\mathrm{int}\nu^{-1}\right)^2\,,
\label{r:size_vs_freq}
\end{equation}
where $\theta_\mathrm{scat}$ and $\theta_\mathrm{int}$ are the scattered and
intrinsic FWHM diameters of the VLBI core, respectively, scaled to a frequency
of 1~GHz. We used a grid-search method to fit for $\theta_\mathrm{scat}$, 
$\theta_\mathrm{int}$, and $k$ that produced the minimum $\chi^2$. We found 
the following best-fitting values $\theta_\mathrm{scat}=55.8$~mas, 
$\theta_\mathrm{int}=4.2$~mas, and the index $k=-1.89$, which corresponds to 
$\beta=4.3$. This indicates that the angular size of the VLBI core increases 
faster than would be expected if there is no scattering, but not as fast as in 
case of scattering caused by a Kolmogorov electron density spectrum.

Partly, this can be due to the non-simultaneity of the VLBI observations, 
given the maximum epoch difference of $\sim$16~yr between the 2/8 GHz and 
15~GHz data. On the other hand, the observations at 2 and 8~GHz at four 
epochs were carried out simultaneously, thereby providing more robust 
estimates of the $k$-index, although determined within a $\sim$10 times 
narrower frequency range, with the average value of $-1.90\pm0.13$ 
($\beta\approx4.2$). This is consistent with the one calculated from 
the all non-simultaneous data covering a broader frequency range. This 
average two-frequency value of $k$ also agrees well with the power index 
$k=-1.97\pm0.29$ derived by \cite{Fey_et_al_89} from earlier non-simultaneous 
VLBI observations of 2023+335 performed at 0.61, 1.66, and 4.99~GHz, but 
still deviates from the conventional index of $-2.2$ expected for a 
Kolmogorov spectrum. The following questions then arise: 
(i)   is it a result of a non-Kolmogorov turbulence, e.g., caused by anisotropy effects?, 
(ii)  what may cause the weaker frequency dependence?, and 
(iii) why the slope of \cite{Fey_et_al_89} ($-1.97$) is closer to that of Kolmogorov 
($-2.2$) than our estimate of the slope ($-1.89$)? 
The currently most accepted model of magnetohydrodynamic 
turbulence developed by \cite{Goldreich95}, in which the compressible regime 
was tested numerically \citep{Cho02,Cho03}, includes anisotropy effects and 
predicts the Kolmogorov spectrum. As for the second question, \cite{Cordes01} 
reported that the scaling law of angular size with frequency can be considerably 
shallower than the canonical $\nu^{-11/5}$ dependence if the scattering screen 
with Kolmogorov fluctuations has a confined structure or is truncated transverse 
to the line of sight, i.e. the ``container'' in which the turbulence resides has 
boundaries. This is consistent with the observed ESE that likely indicates 
the presence of discrete density structures, the observational manifestations of 
which we discussed in Sect.~\ref{s:refrac_scat}. To speculate on the third question, 
we see two possible explanations: (1) the data used by \cite{Fey_et_al_89} probed 
somewhat different (larger) spatial scales, because the lowest frequency in their
data was 0.61~GHz, while that in our set of data is higher by a factor of 2.3, 1.4~GHz;
(2) the properties of the screen, as a function of time, could be different at epoch 
of 1985-1986 studied by \cite{Fey_et_al_89}. Most probably, both of these possibilities 
play a role to some degree.

The most prominent angular broadening occurs at the lowest of the available 
observing frequencies, 1.4 GHz, and increases the angular size by a factor 
of $\sim$10 as seen in Fig. \ref{f:size_vs_freq}. The expected size of the 
core component, if there were no screen, is within typical VLBI core sizes 
of $\sim$(1-2)~mas at 2~GHz and $\sim$(0.3-0.6)~mas at 8~GHz as derived 
from a sample of 370 sources \citep{PK2012}, and $\sim$(0.2-0.4)~mas at 
15~GHz from a sample of 133 sources \citep{Kovalev05}.

From the above analysis, we can constrain the turbulence spectrum
index toward 2023+335 to within the range
$4.2\lesssim\beta\lesssim4.7$.  We note that such steep turbulence
spectra ($\beta\sim4.3$) have been seen previously in nature,
specifically in laboratory scattered laser light experiments
\citep{Jakeman84, Walker84}, suggesting that spectra that are steeper
than the Kolmogorov spectrum may not be that unusual.

\section{Summary}
\label{s:summary}

We have found convincing evidence for the first detected multiple imaging
of an AGN jet due to refractive foreground scattering in our
galaxy. This rare phenomenon was theoretically predicted several decades
ago and is based on the refractive properties of localized electron
density enhancements in the ionized component of the Galactic
interstellar medium \citep{Lovelace70,Cordes86b,Rickett88}. We 
detected the effect in the low galactic latitude ($b=-2\fdg4$) quasar 
2023+335 in a number of VLBA observations obtained between 2008 and 
2012 as part of the MOJAVE program.  Its strongest manifestation, 
at $\sim$10\% of the primary image flux density, occurred on May 28, 
2009 at 15~GHz, when the source was undergoing an extreme scattering 
event, specifically during a special phase when a caustic spike 
associated with the lens edge passed over the source.  We observed a 
highly significant multi-component pattern of secondary images 
induced by strong refraction in May 2009 and July 2009, which was 
stretched out roughly along the constant galactic latitude line with
$\mathrm{PA}\approx40\degr$. This suggests that the direction of
relative motion of the lens is parallel to the galactic plane, as
expected for an orbiting cloud.  We observed sporadic weaker but still
detectable patterns at similar position angles with respect to the
VLBI core at other epochs. Using archival VLBA data, we were able to
detect the same effect at a lower observing frequency, 8.4~GHz. The
angular separation of the peaks of the scatter-induced structure
follows a $\lambda^2$ dependence, which provides strong evidence for a
plasma scattering origin of the multiple imaging. The parsec-scale
source structure, when unaffected by scattering, consists of a bright
core that typically accounts for $\sim$80\% of the total VLBA flux
density, and a short jet extending along $\mathrm{PA}\approx-20\degr$.
Our multiple imaging observations provide valuable information for
discriminating between various competing models of ESE, the true nature 
of which is still uncertain.  

By taking into account the length of the ESE event ($\sim$0.14~yr) and
by assuming that it was caused by an ionized gas cloud drifting across
the line of sight, we determined the proper motion of the lens:
$\sim$6.8~mas~yr$^{-1}$, and its angular size: $\sim$0.27~mas. The
latter is comparable to the intrinsic size of the VLBI core at 15~GHz:
$\sim$0.28~mas ($\sim$1~pc). The line of sight toward 2023+335 passes
through the highly turbulent Cygnus region located at a distance of
about 1.5~kpc. Assuming that that the lens lies at this distance, its
transverse speed and linear size are $\sim$56~km~s$^{-1}$ and
$\sim$0.4~AU, respectively.

Our observations support a model of ESE suggested by
\cite{Clegg98}, since formation of multiple imaging of the
background source along with formation of caustics surfaces in a
light curve predicted by the model have been detected. However, we
were not able to detect any substantial angular position wander of
the background source, as predicted by the model.

Analyzing the non-simultaneous multi-frequency VLBI observations
covering a frequency range from 1.4 to 86~GHz, we found that the
scattered angular size of the VLBI core scales as $\nu^{-1.89}$,
implying the presence of a highly turbulent, refractive dominated
scattering screen along the line of sight to 2023+335. The shallower 
than canonical Kolmogorov $\nu^{-2.2}$ dependence may be caused by a 
confined structure of the screen with arbitrary spatial variations 
of scattering power in the transverse direction.

We note that future multiple imaging events in AGN 
caused by refraction scattering in the interstellar medium can be 
successfully detected if the following two conditions are fulfilled: 
(i) a target is observed by a multi-band VLBI when it is passing through 
the caustic surface of an ESE; (ii) the refractive power of the lens is 
sufficiently strong to separate the secondary images at angular distances 
larger than the scattered diameter of the primary image.

\begin{acknowledgements}

The authors acknowledge K.~I.~Kellermann, E.~Clausen-Brown, and the other 
members of the MOJAVE team. We are also grateful to J.~Cordes, B.~Rickett
and T.~J.~W.~Lazio 
for productive discussions. We thank the anonymous referee for useful comments 
which helped to improve the manuscript. This research has made use of data 
from the MOJAVE database that is maintained by the MOJAVE team \citep{MOJAVE}. 
The MOJAVE project is supported under NASA {\it Fermi} grant 11-Fermi11-0019. 
Part of this work was supported by the COST Action MP0905 ``Black Holes in a 
Violent Universe''. A.~B.~P.\ was partially supported by DAAD and the 
``Non-stationary processes in the Universe'' Program of the Presidium of the 
Russian Academy of Sciences. Y.~Y.~K.\ was supported by the Russian Foundation 
for Basic Research (projects 11-02-00368, 12-02-33101), the basic research program 
``Active processes in galactic and extragalactic objects'' of the Physical Sciences 
Division of the Russian Academy of Sciences, and the Dynasty Foundation. T.~H.\ was 
supported by the Jenny and Antti Wihuri foundation. E.~R.\ was partially supported 
by the Spanish MINECO projects AYA2009-13036-C02-02 and AYA2012-38491-C02-01 and by 
the Generalitat Valenciana project PROMETEO/2009/104. The UMRAO monitoring program is 
supported in part by NASA {\it Fermi} GI grants NNX09AU16G, NNX10AP16G and NNX11AO13G 
and NSF grant AST-0607523. The OVRO 40-m monitoring program is supported in part by 
NASA grants NNX08AW31G and NNX11A043G, and NSF grants AST-0808050 and AST-1109911. 
The VLBA is a facility of the National Science Foundation operated by the National 
Radio Astronomy Observatory under cooperative agreement with Associated Universities, 
Inc. This research has made use of BM046, RDV26,40, BM290A-E, BR149, BM167, BM231, 
GM064, BR145D experiment data from the public VLBA archive of correlated data.
This work made use of the Swinburne University of Technology software correlator 
\citep{Deller11}, developed as part of the Australian Major National Research 
Facilities Programme and operated under licence.

\end{acknowledgements}

\bibliographystyle{aa}
\bibliography{pushkarev}

\end{document}